\documentclass[mathleft
]{an}
\usepackage{graphicx}
\usepackage{times}
\begin{document}

\Pagespan{789}{}
\Yearpublication{2006}%
\Yearsubmission{2005}%
\Month{11}%
\Volume{999}%
\Issue{88}%

\title{High energy emission from AGN cocoons in clusters of galaxies}

\author{M. Kino\inst{1,2}\fnmsep\thanks{Corresponding author:
  \email{motoki.kino@nao.ac.jp}\newline}
 N. Kawakatu\inst{2},
 H. Ito\inst{3}
\and  H. Nagai\inst{2}
}
\titlerunning{High energy emission from cocoons}
\authorrunning{M. Kino, N. Kawakatu, H. Ito \& H. Nagai}
\institute{
ISAS/JAXA, 3-1-1 Yoshinodai, 229-8510 Sagamihara, Japan
\and 
National Astronomical Observatory of Japan, 181-8588 Mitaka, Japan
\and 
Department of Science and Engineering, Waseda University, Tokyo 169-8555,
Japan}

\received{30 May 2005}
\accepted{11 Nov 2005}
\publonline{later}


\abstract{%
Gamma-ray emission from cocoons of young 
radio galaxies is predicted. 
Considering the process of adiabatic injection of 
the shock dissipation energy and mass of the relativistic 
jet into the cocoon, we find that the thermal 
electron temperature of the cocoon is typically predicted 
to be of the order of $\sim$ MeV, and is determined only 
by the bulk Lorentz factor 
of the jet. Together with the time-dependent 
dynamics of the cocoon expansion, we find that young cocoons 
can yield thermal Bremsstrahlung emissions at energies $\sim$MeV. 
Hotter cocoons (i.e., GeV) for younger sources
are also discussed.}

\maketitle

\section{Introduction}

Relativistic jets in active galactic 
nuclei (AGNs) are widely believed to be 
the dissipation of kinetic energy of 
relativistic motion with a Lorentz
factor of order $\sim 10$ produced at the 
vicinity of a super-massive black hole at the
galactic center (Begelman, Blandford and Rees 1984 for reviews).
The jet in
powerful radio loud AGNs (i.e., FR II radio sources)
is slowed down via strong terminal shocks which are identified as 
hot spots.
The shocked plasma then expand sideways and 
envelope the whole jet system and this is so called a cocoon or a bubble
(Fig. 1).
The cocoon is a by-product of the interaction
between AGN jets and surrounding intra-cluster medium (ICM).
The internal energy of the shocked plasma continuously 
inflates this cocoon. 
So far little attention has been paid to
observational feature of the cocoon, since 
they are usually invisible in GHz bands
because of the synchrotron cooling for older electrons.
\footnote{On the contrary,
the emission from the shell made of the shocked ICM (see Fig. 1)
has been explored by many authors
(e.g., Heinz, Reynolds and Begelman 1998; Sutherland and Bicknell 2007).
Since the shells have non-relativistic velocities, 
the emission from the shells is predicted in the X-ray band.} 
As a result, we just see a part of the cocoon.
The visible part is so-called radio lobes in which 
relatively fresh electrons are filled in.
In Fig. 2, 
we show one good sample of cocoon emission from the 
powerful radio galaxy Cygnus A for the evidence of its 
existence.
\footnote{The data obtained by Carilli et al. (1991) was
 re-analyzed to obtain the map shown in Fig. 2.  The observation
 was carried out with VLA A configuration at 330~MHz on 1987 August 18. 
 The data analyzes is performed by standard manner with Astronomical Image
 Processing System (AIPS).  The flux-scale is determined by comparison with 
3C~286 and 3C~48 using the AIPS task SETJY.  The image is obtained with the 
DIFMAP after a number of self-calibration iterations.}

Among a variety of AGN bubbles,
a population so called compact symmetric objects (CSOs) 
has been widely investigated in  various ways
 (e.g., 
 Fanti et al. 1995; 
Readhead et al. 1996;
O'Dea \& Baum 1997;
de Vries et al. 1997;
O'Dea and Baum 1998;
Stanghellini et al. 1998; 
Snellen et al. 2000;
Dallacasa et al. 2000; 
 Giroletti et al. 2003;
Nagai et al. 2006;
Orienti et al. 2007;
Kawakatu et al. 2008).
CSOs are smaller than 1 kpc and the previous studies
support the youth scenario in which CSOs propagate 
from pc scales 
thrusting away an ambient medium and growing up to 
FR II radio galaxies.
CSOs are thus recognized as newly born AGN jets,
and they are crucial sources 
to explore physics of AGN bubbles in their early days.

In this study, 
we propose that ``young AGN bubbles''
are a new population of $\gamma$-ray emitters in the Universe.
The layout of the paper is as follows.
We review the expanding cocoon model in \S2
following our previous works 
(Kino and Kawakatu 2005; Kino, Kawakatu and Ito 2007, KKI07 hereafter)
We show the predicted MeV gamma emission from young cocoons in \S3 .
In \S4, we further predict GeV gamma emission for smaller CSOs.
Summary and discussion is given in \S5.

\section{Cocoon inflation by exhausted jet}

Here we consider the time-evolution of an expanding cocoon  
inflated by the dissipation energy of 
the relativistic jet via terminal shocks.
The adiabatic
energy injection into the cocoon is assumed.
%
Mass and energy conservation from the  jet into
the cocoon, which govern 
the cocoon pressure $P_{\rm c}$ and 
mass density $\rho_{\rm c}$ 
are written as
\begin{eqnarray}\label{eq:pc}
\frac{{\hat \gamma}_{c}}{{\hat \gamma}_{c}-1}
\frac{P_{\rm c}(t)V_{c}(t)}{t}\approx
2 T^{01}_{\rm j}(t)  
A_{\rm j}(t) 
\end{eqnarray}
\begin{eqnarray}\label{eq:rho}
\frac{\rho_{\rm c}(t)V_{c}(t)}{t }\approx
2 J_{\rm j}(t)  
A_{\rm j}(t)   ,
\end{eqnarray}
where
${\hat \gamma}_{c}$,
$V_{\rm c}$,
$T^{01}_{\rm j}$, 
$J_{\rm j}$ and
$A_{\rm j}$,
are
the adiabatic index of the plasma in the cocoon,
the volume of the cocoon,
the kinetic energy and mass flux of the jet, and 
the cross-sectional area of the jet,
respectively.
The total kinetic energy and mass flux of the jet are
$T_{\rm j}^{01}=\rho_{\rm j}c^{2}\Gamma_{\rm j}^{2}v_{\rm j}$,
$J_{\rm j}=\rho_{\rm j}\Gamma_{\rm j}v_{\rm j}$ where
$\rho_{\rm j}$, and 
$\Gamma_{\rm j}$ are
mass density and bulk Lorentz factor of the jet
(Blandford and Rees 1974). 
Hereafter we set $v_{\rm j}=c$.
The total kinetic power of the relativistic jet 
is defined as 
$L_{\rm j}\equiv 2 T^{01}_{\rm j}(t) A_{\rm j}(t)$ 
and it is assumed to be constant in time.
\begin{figure}
\centering
\includegraphics[width=65mm]{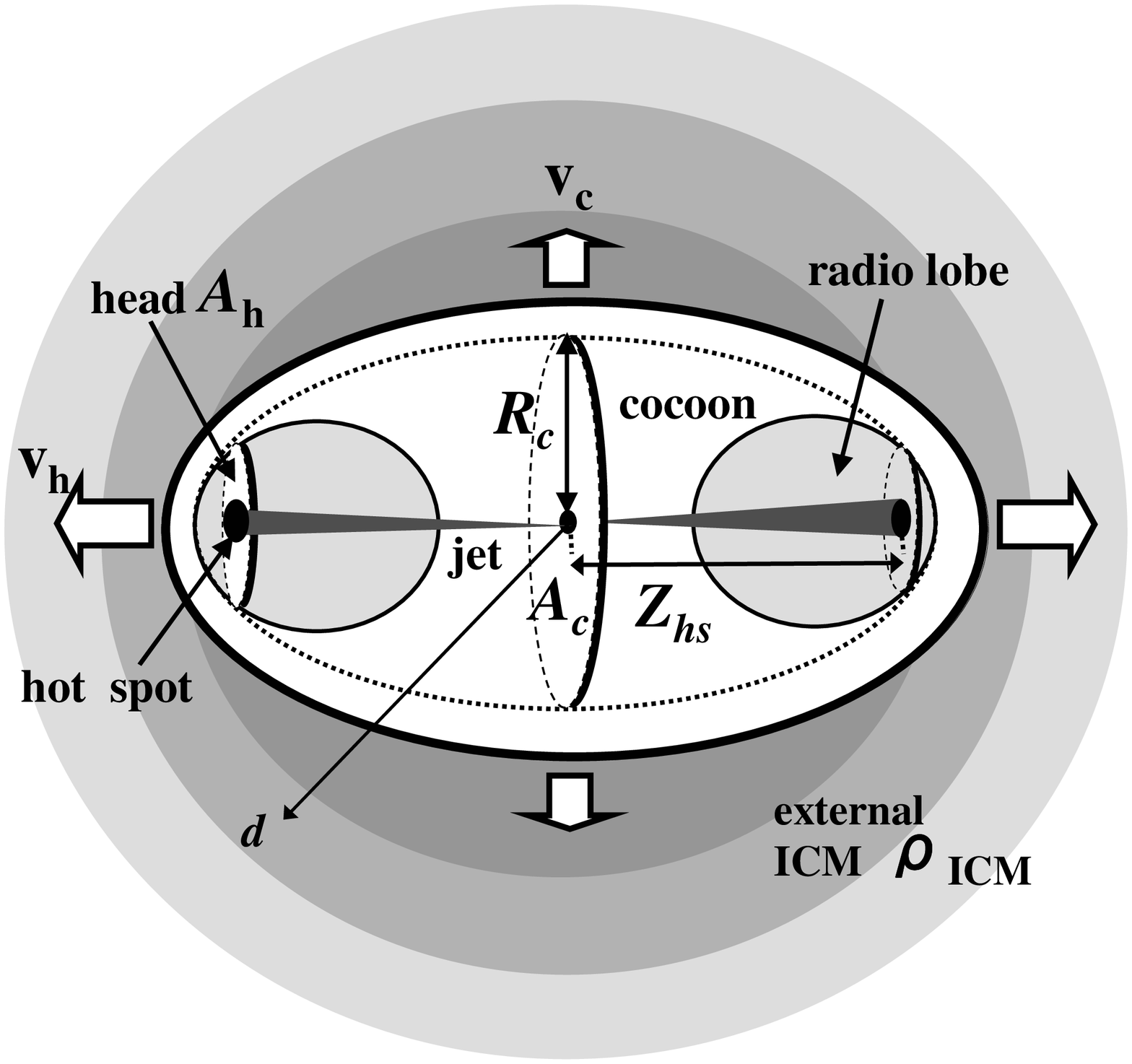}
\caption{A cartoon representation of interaction of 
the ICM with declining atmosphere  
and the relativistic jet in FR II radio galaxy. 
As a result, most of the kinetic energy
of jet is deposited in the cocoon and it is 
inflated by its internal energy.}
\label{label1}
\end{figure}
\begin{figure}
\includegraphics
[width=80mm]
{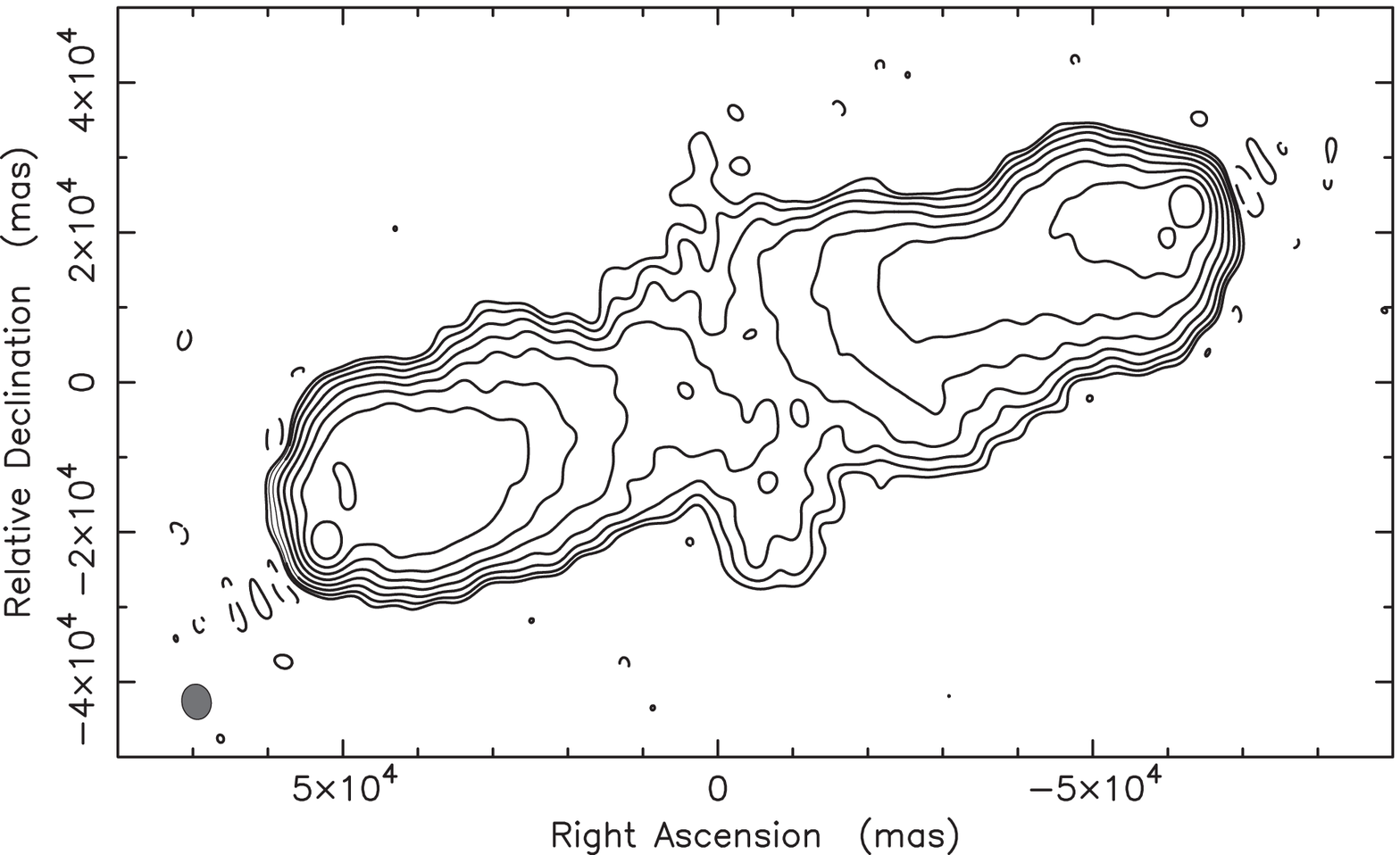}
\caption{
VLA image of Cygnus A at 330~MHz (A-configuration).  
The contour level starts with 0.565~Jy~beam$^{-1}$ and increases 
from there by factors of 2.  The convolving beam is $4.73\times3.91$~arcsec 
at the position angle of $10.1^{\circ}$.
In this frequency, we can see the synchrotron emission
from slightly colder electrons than those emitting GHz ranges.
Hence we can identify
the cocoon profile which are clearly different from familiar
``classical double lobe".}
\label{label1}
\end{figure}

As for the mass and kinetic energy flux of 
powerful relativistic jets,
numerical simulations tell us that
no significant entrainment of the environmental matter
takes place during the jet propagation (e.g., Mizuta et al. 2004).
According to this, 
the mass and kinetic energy flux of the jet are 
regarded as constant in time.
Then, the conditions of $T_{\rm j}^{01}={\rm const}$, and
$J_{\rm j}={\rm const}$ leads to the relations of  
$\rho_{\rm j}(t)A_{\rm j}(t)={\rm const}$ and
$\Gamma_{\rm j}(t)={\rm const}$.
In order to evaluate $L_{\rm j}$,  we use
the shock jump condition of 
$\Gamma_{\rm j}^{2}\rho_{\rm j} =\beta_{\rm hs}^{2}
\rho_{\rm ICM}$ (Kawakatu and Kino 2006)
where 
$\beta_{\rm hs}(=v_{\rm hs}/c)$ and
$\rho_{\rm ICM}$ is the  
advance speed of the  hot spot $\beta_{\rm hs}= 10^{-2}\beta_{-2}$ and
the mass density of ICM, respectively.
Using, the jump condition, 
$L_{\rm j}$ is given by 
\begin{eqnarray}
L_{\rm j}
=2 \times 10^{45}~
R_{\rm kpc}^{2}
\beta_{-2}^{2}
n_{-2}
~{\rm erg \ s^{-1}}
\end{eqnarray}
where
we use $A_{\rm j}(t)=\pi R_{\rm hs}^{2}(t)$, and 
the hot spot radius $R_{\rm hs}$ is given by
$R_{\rm kpc}=R_{\rm hs}(10^{7}~{\rm yr})/1~{\rm kpc}$.
As a fiducial case, we set
the number density of the surrounding ICM as  
$n_{\rm ICM}(d)=\rho_{\rm ICM}(d)/m_{p}= 10^{-2}~{\rm cm^{-3}} 
n_{-2} (d/30~{\rm kpc})^{-2}$ 
where $d$ is the distance from
the center of ICM and cocoon (see Fig. 1).
Since the change of the index from $-2$
does not change the essential 
physics discussed in this work,
we focus on this case for simplicity.
Since $L_{\rm j}$ is the ultimate source of the
phenomena associated with the cocoon, all of the 
emission powers which will appear in \S 3 
should be less than $L_{\rm j}$.

The number density of total electrons in
the cocoon  is 
governed by the cocoon geometry and its plasma content.
For convenience, we define
the ratio of 
``the volume swept by the unshocked relativistic jet"
to ``the volume of the cocoon''
as ${\cal A}(t)$.
We denote 
$V_{\rm c}(t)=2(\pi/3){\cal R}^{2}Z_{\rm hs}^{3}(t)$,
$Z_{\rm hs}$ satisfies 
$Z_{\rm hs}(t)=\beta_{\rm hs}ct$, 
$R_{\rm c}$, and 
${\cal R}\equiv R_{\rm c}/Z_{\rm hs}<1$
as the cocoon volume,
the distance from the central engine to the hot spot, 
is the radius of the cocoon body, and 
the aspect-ratio of the cocoon, respectively
(e.g., Kino and Kawakatu 2005).
Postulating that ${\cal R}$ and 
$Z_{\rm hs}/R_{\rm hs}$ are constant in time, 
${\cal A}(t)\equiv
\frac{2A_{\rm j}(t)v_{\rm j} t}{V_{\rm c}(t)}$ is evaluated as
\begin{eqnarray}
{\cal A}(t)
\approx
0.4~
{\cal R}^{-2}
R_{\rm kpc}^{2}
Z_{30}^{-2}
\beta_{-2}^{-1}
\end{eqnarray}
where
$Z_{30}= 
Z_{\rm hs}(10^{7}~{\rm yr})
/30~{\rm kpc}$.
Note that, in the case,
the time dependence of ${\cal A}$ is deleted
since $V_{\rm c}\propto t^{-3}$ and $A_{\rm j}\propto t^{2}$.
This case satisfies  $v_{\rm hs}={\rm const}$
(e.g., Conway 2002).
The cocoon mass density $\rho_{c}(t)$
is controlled by the mass injection by the jet and 
it can be expressed as
$\rho_{\rm c}(t)
\approx
\Gamma_{\rm j}
\rho_{\rm j}(t)
{\cal A}  
=
\beta_{\rm hs}^{2}
\Gamma_{\rm j}^{-1}
\rho_{\rm ICM}(Z_{\rm hs}(t))
{\cal A}$
where we use the shock condition of 
$\Gamma_{\rm j}^{2}\rho_{\rm j} =\beta_{\rm hs}^{2}
\rho_{\rm ICM}$. 
Adopting typical quantities of FR II sources
(e.g., Begelman, Blandford and Rees 1984),
the number density of the total electrons in the cocoon is 
given by
\begin{eqnarray} \label{eq:ne}
n_{e}(t)
\approx
4 \times 10^{-5} 
\bar {{\cal A}}
n_{-2}
\Gamma_{10}
\beta_{-2}^{2}
\left(\frac{t}{10^{7}~{\rm yr}}
\right)^{-2}
{\rm cm^{-3}}
\end{eqnarray}
where
$\Gamma=10 \Gamma_{10}$, and
$\bar{{\cal A}}={\cal A}/0.4$. 
Here
we assume that the mass density of 
the $e^{\pm}$ pair plasma is heavier
than that of electron-proton one,
and then we adopt  $\rho_{\rm c}\approx 2 m_{e}n_{e}$
in the light of previous works 
(Reynolds et al. 1996; 
Wardle et al. 1998;
Sikora and Madejski 2000;
Kino and Takahara 2004). 
The upper limit of  thermal $n_{e}$ can be basically constrained by 
the analysis of Faraday depolarization (Dreher et al. 1987).
However, the strong Faraday
depolarization observed in CSOs (Cotton et al. 2003)
are likely to be caused by
dense foreground matter such as narrow line region.
Therefore  $n_{e}$ in radio lobes of CSOs 
has not been clearly constrained.

Let us estimate
the electron (and positron) temperature ($T_{e}$)
and proton temperature ($T_{p}$).
From Eqs. (\ref{eq:pc}) and (\ref{eq:rho})
together with the equation of state 
$P_{\rm c}\approx  2n_{e}kT_{e}$,
we can directly derive the temperatures as
\begin{eqnarray} \label{eq:Te}
kT_{e} \approx 1~ \Gamma_{10} ~ {\rm MeV}, \quad
kT_{p} \approx 2 ~ \Gamma_{10} ~ {\rm GeV}
\end{eqnarray}
where we adopt the two temperatures condition of
$kT_{e}\approx (m_{e}/m_{p}) k T_{p}$.
It should be stressed that 
the temperatures are governed only by $\Gamma_{\rm j}$.
It is  also worth noting that the geometrical
factors in Eqs. (\ref{eq:pc}) and (\ref{eq:rho}) are completely
cancelled out.
One can naturally understand these properties
by comparing the well-established properties 
such as supernovae and GRBs.
Constant temperature in AGN jet can be realized by 
the ``continuous'' energy 
injection into the expanding cocoon 
whilst temperatures of 
astrophysical explosive sources such as gamma-ray bursts
and supernovae would be decreased
because of ``impulsive'' injection of the energy. 
Thus the resultant temperatures are
uniquely governed by $\Gamma_{\rm j}$ and they remain to be 
constant in time.

\section{MeV $\gamma$ emissions from a young cocoon}
\subsection{Thermal MeV Bremsstrahlung emission}

The time-dependence of 
the thermal Bremsstrahlung luminosity
$L_{\rm Brem}$ is given by $L_{\rm Brem}(t)\propto
n_{e}^{2}(t)T_{e}^{3/2}V_{\rm c}(t)
\propto t^{-1}$ based on the cocoon expansion
shown in the previous section.
Hence it is clear that a younger cocoons are 
brighter Bremsstrahlung emitters than older cocoons.
In a similar way,
brighter synchrotron luminosity has been expected
for younger radio galaxies (e.g., Readhead et al. 1996).
With relativistic thermal
Bremsstrahlung emissivity (Rybicki and Lightman 1979), 
the luminosity of the 
optically thin thermal Bremsstrahlung emission
$\nu L_{\nu}$ at energies $\sim 1~{\rm MeV}$ 
is estimated as 
\begin{eqnarray}\label{eq:l_brem}
L_{\rm Brem}(t)\approx 2\times 10^{40}~
\bar{n}_{e}^{2}
{\cal R}^{2}
\Theta_{10}^{3/2}
\left(\frac{t}{10^{7}~{\rm yr}}\right)^{-1}~{\rm erg \ s^{-1}} .
\end{eqnarray}
Eq.  (\ref{eq:l_brem}) explains the reasons  for 
the non-detection of the thermal emission from  
older cocoons.
One is simply because it is not very bright.
The other is because the predicted energy range is $\sim 1~{\rm MeV}$,
the MeV-$\gamma$ astronomy is still immature and it is sometimes 
called as ``sensitivity gap'' compared with 
the energy range below 10 keV 
and above GeV ranges (Takahashi et al. 2004).
In Fig. 3, we show 
the predicted values of $\nu F_{\nu}$
for the cocoons with $t=10^{7}~{\rm yr}$ and 
$t=10^{4}~{\rm yr}$
located at the  distance of $D=10^{2}~{\rm Mpc}$.
The cocoon with $t=10^{7}~{\rm yr}$
have $\nu F_{\nu}\sim 10^{-14}~{\rm erg~cm^{-2}s^{-1}}$.
The detection threshold of 
SPI instrument on board the INTEGRAL satellite is about
$\nu F_{\nu}\sim 10^{-9}~{\rm erg~cm^{-2}s^{-1}}$ 
at $\sim 1~{\rm MeV}$.
For a young cocoon with $t=10^{4}~{\rm yr}$, the 
predicted luminosity is 
$\sim 10^{3}$ times larger than that 
$\nu F_{\nu}\sim 10^{-11}~{\rm erg~cm^{-2}s^{-1}}$. 
This is still less than the threshold of 
INTEGRAL. This may be the reason for the lack of
detection of MeV emission from young cocoons up to now.
Fig. 3 shows that 
the XMM/Newton satellites can detect the 
low energy part of the thermal Bremsstrahlung from young cocoons.
In MeV energy band, a proposed mission of detector
SGD on board the NeXT satellite 
with the eye up to $\sim 0.6~{\rm MeV}$ (Takahashi et al. 2004)
could detect the thermal MeV emission 
from those located slightly 
closer or younger with smaller Lorentz factor.

\subsection{Candidate sources}

In order to explore the extended cocoon  emission
in the X-ray band, one may think it is hard to distinguish 
overlapping emission from the compact core of the AGN
with limited spacial angular resolution
of the current X-ray satellites.
However, the averaged spectral index  
in X-ray band ($\Gamma_{X}$) from 
the compact core  of AGNs is  softer
than the Bremsstrahlung emission (Koratkar and Blaes 1999). 
Hence it is possible find candidate sources 
of the MeV cocoon by the value of $\Gamma_{X}$.
As far as we know, there are two possible candidates for
the Bremsstrahlung emission.
Those are  B1358+624 (Vink at al. 2006) 
and PKS B1345+125 (Siemiginowska et al. 2008) actually 
shows $\Gamma_{X}\approx 1$,
both of them were observed by XMM/Newton.
Time variability of 
observed spectra is also the key to distinguish them. 
It is obvious that the cocoon emission is 
constant in time whilst various emissions
from the core of AGN should be highly variable.
Hence steady emissions are
convincingly originated in cocoons.

Intriguingly,
``the diffuse X-ray emission'' has been indeed detected
in PKS B1345+125 with 
the size of the extended emission is of order $\sim 20~{\rm kpc}$
by Siemiginowska et al. (2008).
The diffuse emission might  
be associated with the radio lobes of this source.
The X-ray emission is elongated towards
the South-West similarly to the VLBI jet axis 
reported by Stanghellini et al. (2001).
If the emission is associated with the radio emitting 
plasma, there are two possibilities to explain the emission.
Here we newly stress that 
the tail of the Bremsstrahlung emission 
could explain the emission.
Based on Eq. (\ref{eq:l_brem}), the X-ray luminosity of the emission shows
$L_{X}\approx 1\times 10^{43}~{\rm erg~s^{-1}}$.
The observed $L_{X}$ can be explained with the
$n_{e}\approx 1\times 10^{-2}~{\rm cm^{-3}}$.
Based on Eq. (\ref{eq:ne}), the $n_{e}$ can be realized 
with $\beta_{\rm hs}\sim 10^{-1}$ and 
$n_{\rm ICM}\sim 10^{-1}~{\rm cm^{-3}}$ for instance.
Since the required $n_{e}$ is considerably large,
the analysis of Faraday depolarization will be crucial 
for checking the upper limit of  $n_{e}$.
The other possibility is non-thermal emissions 
from the lobes which is recently investigated by
Stawarz et al. (2008).
Observations near the peak of thermal emission 
NeXT satellite will be crucial to distinguish whether 
the emission is thermal or non thermal one.

\section{Younger radio sources as ``hotter" bubbles}

So far,  
we discuss the cocoon
property in the phase 
of no significant 
cooling.
The phase  roughly corresponds to 
medium size symmetric objects 
(MSOs) 
and FR II galaxies.
Since cooling timescales become shorter for smaller sources,
cooling effects 
for CSOs are 
more effective than 
the case for larger ones 
such 
as MSOs 
and 
FR IIs. We consistently solve a set of equations
which describes young bubble expansions
including the effects of Bremsstrahlung emission
and adiabatic loss 
together with the initial conditions indicated by CSO observations.
Then we find that the bubbles have electron temperature of $\sim$ GeV 
at initial phases,
the bubbles then cooled down to MeV by the adiabatic loss.
We further estimate these $\gamma$-ray emissions and
show that it could be detected with Fermi (GLAST)
(Kino et al. 2009).

\section{Summary and discussion}

We have investigated the luminosity evolutions 
of AGN cocoons together with the dynamical 
evolution of expanding cocoon.
Below we summarize the main results of the present work.

\begin{enumerate}

\item
We newly predict the Bremsstrahlung 
emission peaked at MeV-$\gamma$ band as a result
of standard shock dissipation
of relativistic jets in AGNs. 
The temperature of the cocoon is 
governed only by the bulk Lorentz factor of the jet $\Gamma_{\rm j}$. 
The electron temperature $T_{e}$ 
relevant to observed emissions
is typically predicted in the range of  MeV
for $\Gamma_{\rm j}\sim 10$.
Constant temperatures of plasma in the cocoon
can be realized because of the continuous energy injection
by the jet with constant $\Gamma_{\rm j}$ (KKI07).

\item

We further investigate  younger bubble expansions
including the effects of Bremsstrahlung emission
and adiabatic loss 
together with the initial conditions of CSOs.
Then we find that the lobes initially 
have electron temperature of GeV and
the lobes then cool down to MeV by the adiabatic loss.
The $\gamma$-ray emissions  could be detected with Fermi (GLAST)
(Kino et al. 2009).

\end{enumerate}

\begin{figure}
\includegraphics
[width=80mm]
{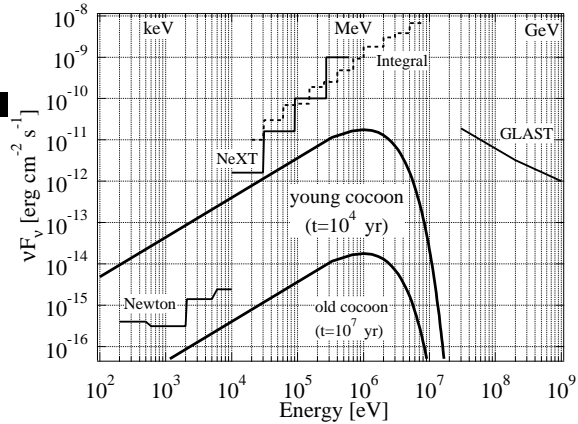}
\caption{
Model prediction of MeV-peaked thermal bremsstrahlung
emission from cocoons located at $D=10^{2}$ Mpc. 
The predicted emission from 
young cocoon is brighter enough to detect in X-ray band
whilst that from an old cocoon is much darker than 
the detection limits.}
\label{label1}
\end{figure}

\acknowledgements

We would like to thank 
C. R. Kaiser and 
M. Sikora 
for valuable comments.
NK is supported by Grant-in-Aid for JSPS Fellows.
HI acknowledge
the Grant for Special Research Projects at Waseda University.

%
%

\end{document}